\documentclass[a4paper,twoside,10pt]{article}

\input{jgrg21.sty}
\usepackage{graphicx}
\usepackage{dcolumn}
\usepackage{amsfonts}
\usepackage{latexsym}
\usepackage{amssymb} 
\usepackage{verbatim}
\usepackage{amsthm}
\usepackage{amsmath}
\newtheorem{conjecture}{Conjecture}
\begin{document}
%
\pagestyle{fancy}
\fancyhead{}
  \fancyhead[RO,LE]{\thepage}
  \fancyhead[LO]{K.~Nakamura}                  
  \fancyhead[RE]{Short note on construction of gauge-invariant variables ...}
\rfoot{}
\cfoot{}
\lfoot{}
\label{O28-09}    
\title{%
  Short note on construction of gauge-invariant variables of
  linear metric perturbations on an arbitrary background
  spacetime
}
%
\author{%
  Kouji Nakamura\footnote{Email address: kouji.nakamura@nao.ac.jp}
}
%
\address{%
  TAMA Project Office, Optical and Infrared Astronomy Division,\\
  National Astronomical Observatory of Japan,\\
  2-21-1, Osawa, Mitaka, Tokyo 181-8588, Japan  
}
%
\abstract{
  An outline of a proof of the decomposition of linear metric
  perturbations into gauge-invariant and gauge-variant parts on
  an arbitrary background spacetime which admits ADM
  decomposition is briefly discussed.
  We explicitly construct the gauge-invariant and gauge-variant
  parts of the linear metric perturbations based on some
  assumptions.
  This implies that we can develop the higher-order
  gauge-invariant perturbation theory on an arbitrary background
  spacetime.
}



{\it Introduction ---} 
Higher-order general-relativistic perturbations have very wide
applications.
Among these applications, second-order cosmological
perturbations are topical subject due to the precise
measurements in recent cosmology.
Higher-order black hole perturbations are also discussed in some
literature.
Moreover, as a special example of higher-order perturbation
theory, there are researches on perturbations of a spherical
star, which are motivated by researches into the oscillatory
behaviors of a rotating neutron star.
Thus, there are many physical situations to which
general-relativistic higher-order perturbation theory should be
applied.


As well-known, general relativity is based on the concept of
general covariance.
Due to this general covariance, the ``gauge degree of freedom'', 
which is unphysical degree of freedom of perturbations, arises
in general-relativistic perturbations.
To obtain physical results, we have to fix this gauge degrees
of freedom or to extract some invariant quantities of
perturbations.
This situation becomes more complicated in higher-order
perturbation theory.
Therefore, it is worthwhile to investigate higher-order
gauge-invariant perturbation theory from a general point of
view.


According to these motivations, in
Ref.~\cite{kouchan-gauge-inv-O28-09}, we proposed a procedure to
find gauge-invariant variables for higher-order perturbations on
an arbitrary background spacetime.
This proposal is based on the single assumption that {\it we
  already know the procedure to find gauge-invariant variables
  for linear-order metric perturbations}
(Conjecture \ref{conjecture:decomposition-conjecture-O28-09} in
this article).
Under this assumption, we summarize some formulae for the
second-order perturbations of the curvatures and energy-momentum
tensor for the matter fields in
Refs.~\cite{kouchan-second-O28-09,kouchan-second-cosmo-matter-O28-09}. 
Confirming that the above assumption is correct in the case of
cosmological perturbations, in
Refs.~\cite{kouchan-cosmo-second-O28-09}, the second-order
gauge-invariant cosmological perturbation theory was developed. 
Through these works, we find that our general framework of
higher-order gauge-invariant perturbation theory is well-defined  
except for the above assumption for linear-order metric
perturbations.
Therefore, we proposed the above assumption as a conjecture in
Ref.~\cite{kouchan-second-cosmo-matter-O28-09}.
If this conjecture is true, our general-relativistic
higher-order gauge-invariant perturbation theory is completely
formulated on an arbitrary background spacetime and has very
wide applications.
The main purpose of this article is to give a brief outline of a
proof of this conjecture.
Details of this issue is given in
Ref.~\cite{K.Nakamura-2011-full-paper-O28-09}.




{\it Perturbations in general relativity ---} 
The notion of ``gauge'' in general relativity arise in the
theory due to the general covariance.
There are two kinds of ``gauges'' in general relativity.
These two ``gauges'' are called as the first- and the
second-kind gauges, respectively.
The distinction of these two different notion of ``gauges'' is
an important premise of our arguments.
{\it The first-kind gauge} is a coordinate system on a single
manifold ${\cal M}$.
The coordinate transformation is also called {\it gauge
  transformation of the first kind} in general relativity.
On the other hand, {\it the second-kind gauge} appears in
perturbation theories in a theory with general covariance.
In perturbation theories, we always treat two spacetime
manifolds.
One is the physical spacetime ${\cal M}$ which is our nature
itself and we want to clarify the properties of ${\cal M}$
through perturbations.
Another is the background spacetime ${\cal M}_{0}$ which has
nothing to do with our nature but is prepared by hand for
perturbative analyses.
{\it The gauge choice of the second kind} is the point
identification map ${\cal X}$ $:$ 
${\cal M}_{0}\mapsto{\cal M}$.
We have to note that the correspondence ${\cal X}$ between
points on ${\cal M}_{0}$ and ${\cal M}$ is not unique in the 
perturbation theory with general covariance.
General covariance intuitively means that there is no
preferred coordinate system in the theory.
Due to this general covariance, we have no guiding principle to 
choose the identification map ${\cal X}$.
Actually, as a gauge choice of the second kind, we may choose a
different point identification map ${\cal Y}$ from ${\cal X}$. 
This implies that there is degree of freedom in the gauge choice
of the second kind.
This is {\it the gauge degree of freedom of the second kind} in
perturbation theory.
{\it The gauge transformation of the second kind} is understood
as a change ${\cal X}\rightarrow{\cal Y}$ of the identification
map.


To define perturbations of an arbitrary tensor field $\bar{Q}$,
we have to compare $\bar{Q}$ on the physical spacetime 
${\cal M}_{\lambda}$ with $Q_{0}$ on the background spacetime
${\cal M}_{0}$ through the introduction of the above second-kind 
gauge choice ${\cal X}_{\lambda}$ $:$ ${\cal M}_{0}$
$\rightarrow$ ${\cal M}_{\lambda}$.
The pull-back ${\cal X}_{\lambda}^{*}$, which is induced by the
map ${\cal X}_{\lambda}$, maps a tensor field $\bar{Q}$ on 
${\cal M}_{\lambda}$ to a tensor field 
${\cal X}_{\lambda}^{*}\bar{Q}$ on ${\cal M}_{0}$. 
Once the definition of the pull-back of the gauge choice 
${\cal X}_{\lambda}$ is given, the perturbations of a tensor
field $\bar{Q}$ under the gauge choice ${\cal X}_{\lambda}$ are
simply defined by the evaluation of the Taylor expansion at 
${\cal M}_{0}$:
\begin{equation}
  \label{eq:Bruni-35-O28-09}
  {}^{\cal X}\!Q
  :=
  \left.{\cal X}^{*}_{\lambda}\bar{Q}_{\lambda}\right|_{{\cal M}_{0}} 
  =
  Q_{0}
  + \lambda {}^{(1)}_{\;\cal X}\!Q
  + \frac{1}{2} \lambda^{2} {}^{(2)}_{\;\cal X}\!Q
  + O(\lambda^{3}),
\end{equation}
where ${}^{(1)}_{\;\cal X}\!Q$ and ${}^{(2)}_{\;\cal X}\!Q$ are
the first- and the second-order perturbations of $\bar{Q}$,
respectively.


When we have two different gauge choices ${\cal X}_{\lambda}$
and ${\cal Y}_{\lambda}$, we have two different representations
of the perturbative expansion of the pulled-backed variables 
$\left.{\cal X}^{*}_{\lambda}\bar{Q}_{\lambda}\right|_{{\cal M}_{0}}$
and 
$\left.{\cal Y}^{*}_{\lambda}\bar{Q}_{\lambda}\right|_{{\cal M}_{0}}$.
Although these two representations of the perturbations are
different from each other, these should be equivalent because of
general covariance.
This equivalence is guaranteed by the 
{\it gauge-transformation rules} between two different gauge
choices.
The change of the gauge choice from ${\cal X}_{\lambda}$ to
${\cal Y}_{\lambda}$ is represented by the diffeomorphism 
$\Phi_{\lambda}:=({\cal X}_{\lambda})^{-1}\circ{\cal Y}_{\lambda}$.
This diffeomorphism $\Phi_{\lambda}$ is the map $\Phi_{\lambda}$
$:$ ${\cal M}_{0}$ $\rightarrow$ ${\cal M}_{0}$ for each value
of $\lambda\in{\mathbb R}$ and does change the point
identification.
Therefore, the diffeomorphism $\Phi_{\lambda}$ is regarded as
the gauge transformation $\Phi_{\lambda}$ $:$
${\cal X}_{\lambda}$ $\rightarrow$ ${\cal Y}_{\lambda}$.
The gauge transformation $\Phi_{\lambda}$ induces a pull-back
from the representation ${}^{\cal X}Q_{\lambda}$ of the
perturbed tensor field $Q$ in the gauge choice 
${\cal X}_{\lambda}$ to the representation 
${}^{\cal Y}Q_{\lambda}$ in the gauge choice 
${\cal Y}_{\lambda}$.
Actually, the tensor fields ${}^{\cal X}Q_{\lambda}$ and
${}^{\cal Y}Q_{\lambda}$, which are defined on ${\cal M}_{0}$,
are connected by the linear map $\Phi^{*}_{\lambda}$ as
${}^{\cal Y}Q_{\lambda}=\Phi^{*}_{\lambda} {}^{\cal X}Q_{\lambda}$.
According to generic arguments concerning the Taylor expansion
of the pull-back of tensor fields on the same
manifold~\cite{K.Nakamura:2010yg-O28-09}, we obtain the
order-by-order gauge-transformation rules for the perturbative
variables ${}^{(1)}Q$ and ${}^{(2)}Q$ as
\begin{eqnarray}
  \label{eq:Bruni-47-one-O28-09}
  {}^{(1)}_{\;{\cal Y}}\!Q - {}^{(1)}_{\;{\cal X}}\!Q = 
  {\pounds}_{\xi_{(1)}}Q_{0}, \quad
  {}^{(2)}_{\;\cal Y}\!Q - {}^{(2)}_{\;\cal X}\!Q = 
  2 {\pounds}_{\xi_{(1)}} {}^{(1)}_{\;\cal X}\!Q 
  +\left\{{\pounds}_{\xi_{(2)}}+{\pounds}_{\xi_{(1)}}^{2}\right\} Q_{0}.
\end{eqnarray}
where the vector fields $\xi_{(1)}^{a}$ and $\xi_{(2)}^{a}$ are
the generators of the gauge transformation $\Phi_{\lambda}$.


The notion of gauge invariance considered in this article is the   
{\it order-by-order gauge invariance} proposed in
Ref.~\cite{kouchan-second-cosmo-matter-O28-09}. 
We call the $k$th-order perturbation ${}^{(k)}_{{\cal X}}\!Q$ is
gauge invariant iff ${}^{(k)}_{\;\cal X}\!Q = {}^{(k)}_{\;\cal Y}\!Q$
for any gauge choice ${\cal X}_{\lambda}$ and
${\cal Y}_{\lambda}$. 
Through this concept of the order-by-order gauge invariance, we
can develop the gauge-invariant perturbation theory.


{\it Construction of gauge-invariant variables ---} 
First, we consider the metric perturbation.
The metric $\bar{g}_{ab}$ on ${\cal M}$, which is pulled back to 
${\cal M}_{0}$ using a gauge choice ${\cal X}_{\lambda}$, is
expanded in the form of Eq.~(\ref{eq:Bruni-35-O28-09}): 
${\cal X}^{*}_{\lambda}\bar{g}_{ab}$ $=$ $g_{ab}$ $+$ 
$\lambda {}_{{\cal X}}\!h_{ab}$ $+$ 
$(\lambda^{2}/2){}_{{\cal X}}\!l_{ab}$ $+$ $O^{3}(\lambda)$,
where $g_{ab}$ is the metric on ${\cal M}_{0}$.
Although this expansion of the metric depends entirely on the
gauge choice ${\cal X}_{\lambda}$, henceforth, we do not
explicitly express the index of the gauge choice 
${\cal X}_{\lambda}$ if there is no possibility of confusion. 
Through these setup, in Ref.~\cite{kouchan-gauge-inv-O28-09}, we 
proposed a procedure to construct gauge-invariant variables for
higher-order perturbations.
Our starting point to construct gauge-invariant variables is the
following conjecture for the linear-order metric perturbation
$h_{ab}$ defined by the above metric expansion:
\begin{conjecture}
  \label{conjecture:decomposition-conjecture-O28-09}
  If there is a symmetric tensor field $h_{ab}$ of the second
  rank, whose gauge transformation rule is 
  ${}_{{\cal Y}}\!h_{ab}$ $-$ ${}_{{\cal X}}\!h_{ab}$ $=$
  ${\pounds}_{\xi_{(1)}}g_{ab}$, then there exist a tensor field
  ${\cal H}_{ab}$ and a vector field $X^{a}$ such that $h_{ab}$
  is decomposed as $h_{ab}$ $=:$ ${\cal H}_{ab}$ $+$
  ${\pounds}_{X}g_{ab}$, where ${\cal H}_{ab}$ and $X^{a}$ are
  transformed as ${}_{{\cal Y}}\!{\cal H}_{ab}$ $-$ 
  ${}_{{\cal X}}\!{\cal H}_{ab}$ $=$ $0$, 
  ${}_{\quad{\cal Y}}\!X^{a}$ $-$ ${}_{{\cal X}}\!X^{a}$ $=$
  $\xi^{a}_{(1)}$ under the gauge transformation
  (\ref{eq:Bruni-47-one-O28-09}), respectively.
\end{conjecture}
In this conjecture, ${\cal H}_{ab}$ and $X^{a}$ are 
{\it gauge-invariant} and {\it gauge-variant} parts of the
perturbation $h_{ab}$.
In the case of the perturbation theory on an arbitrary
background spacetime, this conjecture is a highly non-trivial
statement due to the non-trivial curvature of the background
spacetime, though its inverse statement is trivial.






{\it An outline of a proof of Conjecture
  \ref{conjecture:decomposition-conjecture-O28-09} ---}
To give an outline of a proof of Conjecture
\ref{conjecture:decomposition-conjecture-O28-09} on an arbitrary 
background spacetime, we assume that the background spacetimes
admit ADM decomposition.
Therefore, the background spacetime ${\cal M}_{0}$ considered
here is $n+1$-dimensional spacetime which is described by the
direct product ${\mathbb R}\times\Sigma$. 
Here, ${\mathbb R}$ is a time direction and $\Sigma$ is the
spacelike hypersurface ($\dim\Sigma = n$) embedded in
${\cal M}_{0}$.
This means that ${\cal M}_{0}$ is foliated by the one-parameter 
family of spacelike hypersurface $\Sigma(t)$, where
$t\in{\mathbb R}F$ is 
a time function.
Then, the metric on ${\cal M}_{0}$ is described by 
\begin{eqnarray}
  \label{eq:gdb-decomp-dd-minus-main-O28-09}
  g_{ab} &=& - \alpha^{2} (dt)_{a} (dt)_{b}
  + q_{ij}
  (dx^{i} + \beta^{i}dt)_{a}
  (dx^{j} + \beta^{j}dt)_{b},
\end{eqnarray}
where $\alpha$ is the lapse function, $\beta^{i}$ is the
shift vector, and $q_{ab}=q_{ij}(dx^{i})_{a}(dx^{i})_{b}$ is the
metric on $\Sigma(t)$.


Since the ADM decomposition
(\ref{eq:gdb-decomp-dd-minus-main-O28-09}) is a local one, we
may regard that the arguments in this article are restricted to
that for a single patch in ${\cal M}_{0}$ which is covered by
the metric (\ref{eq:gdb-decomp-dd-minus-main-O28-09}).  
Further, we may change the region which is covered by the metric
(\ref{eq:gdb-decomp-dd-minus-main-O28-09}) through the choice of
the lapse function $\alpha$ and the shift vector $\beta^{i}$.
The choice of $\alpha$ and $\beta^{i}$ is regarded as the
first-kind gauge choice, which have nothing to do with the
second-kind gauge.
Since we may regard that the representation 
(\ref{eq:gdb-decomp-dd-minus-main-O28-09}) of the background
metric is that on a single patch in ${\cal M}_{0}$, in general
situation, each $\Sigma$ may have its boundary $\partial\Sigma$.


To prove Conjecture
\ref{conjecture:decomposition-conjecture-O28-09}, we first
consider the components of the metric $h_{ab}$ as $h_{ab}$ $=$
$h_{tt}(dt)_{a}(dt)_{b}$ $+$ $2h_{ti}(dt)_{(a}(dx^{i})_{b)}$ $+$ 
$h_{ij}(dx^{i})_{a}(dx^{j})_{b}$.
Under the gauge-transformation rule ${}_{{\cal Y}}\!h_{ab}$ $-$
${}_{{\cal X}}\!h_{ab}$ $=$ ${\pounds}_{\xi_{(1)}}g_{ab}$, the
components $\{h_{tt}$, $h_{ti}$, $h_{ij}\}$ are transformed as 
\begin{eqnarray}
  {}_{{\cal Y}}h_{tt}
  -
  {}_{{\cal X}}h_{tt}
  &=&
    2 \partial_{t}\xi_{t}
  - \frac{2}{\alpha}\left(
    \partial_{t}\alpha 
    + \beta^{i}D_{i}\alpha 
    - \beta^{j}\beta^{i}K_{ij}
  \right) \xi_{t}
  \nonumber\\
  &&
  - \frac{2}{\alpha} \left(
    \beta^{i}\beta^{k}\beta^{j} K_{kj}
    - \beta^{i} \partial_{t}\alpha
    + \alpha q^{ij} \partial_{t}\beta_{j}
    + \alpha^{2} D^{i}\alpha 
    - \alpha \beta^{k} D^{i} \beta_{k}
    - \beta^{i} \beta^{j} D_{j}\alpha 
  \right)\xi_{i}
  \label{eq:gauge-trans-of-htt-ADM-BG-O28-09}
  , \\
  {}_{{\cal Y}}h_{ti}
  -
  {}_{{\cal X}}h_{ti}
  &=&
  \partial_{t}\xi_{i}
  + D_{i}\xi_{t}
  - \frac{2}{\alpha} \left(
    D_{i}\alpha 
    - \beta^{j}K_{ij}
  \right) \xi_{t}
  - \frac{2}{\alpha} M_{i}^{\;\;j} \xi_{j}
  \label{eq:gauge-trans-of-hti-ADM-BG-O28-09}
  , \\
  {}_{{\cal Y}}h_{ij}
  -
  {}_{{\cal X}}h_{ij}
  &=&
  2 D_{(i}\xi_{j)}
  + \frac{2}{\alpha} K_{ij} \xi_{t}
  - \frac{2}{\alpha} \beta^{k} K_{ij} \xi_{k}
  \label{eq:gauge-trans-of-hij-ADM-BG-O28-09}
  ,
\end{eqnarray}
where $M_{i}^{\;\;j}$ is defined by $M_{i}^{\;\;j}$ $:=$ $-$
$\alpha^{2}K^{j}_{\;\;i}$ $+$ $\beta^{j}\beta^{k} K_{ki}$ $-$
$\beta^{j}D_{i}\alpha$ $+$ $\alpha D_{i}\beta^{j}$. 
Here, $K_{ij}$ is the components of the extrinsic curvature of 
$\Sigma$ in ${\cal M}_{0}$ and $D_{i}$ is the covariant
derivative associate with the metric $q_{ij}$ ($D_{i}q_{jk}=0$). 
The extrinsic curvature $K_{ij}$ and its trace $K$ are related
to the time derivative of the metric $q_{ij}$ by  
$K_{ij}$ $=$ $-$ $(1/2\alpha)$
$\left[\partial_{t}q_{ij}-D_{i}\beta_{j}-D_{j}\beta_{i}\right]$
and $K$ $:=$ $q^{ij}K_{ij}$, respectively.


Inspecting gauge-transformation rules
(\ref{eq:gauge-trans-of-htt-ADM-BG-O28-09})--(\ref{eq:gauge-trans-of-hij-ADM-BG-O28-09}),
we consider the decomposition of the components
$\{h_{ti},h_{j}\}$ into the set of the variables
$\{h_{(VL)}$, $h_{(V)i}$, $h_{(L)}$, $h_{(TV)i}$, $h_{(TT)ij}\}$
as follows: 
\begin{eqnarray}
  &&
  h_{ti}
  =:
  D_{i}h_{(VL)} + h_{(V)i}
  - \frac{2}{\alpha} \left(
    D_{i}\alpha 
    - \beta^{k}K_{ik}
  \right) \left(
    h_{(VL)}
    - \Delta^{-1}D^{k}\partial_{t}h_{(TV)k}
  \right)
  - \frac{2}{\alpha} M_{i}^{\;\;k} h_{(TV)k}
  \label{eq:hti-decomp-alternative-O28-09}
  , \\
  &&
  h_{ij}
  =:
  \frac{1}{n} q_{ij} h_{(L)} +
  D_{i}h_{(TV)j}+D_{j}h_{(TV)i}-\frac{2}{n}q_{ij}D^{k}h_{(TV)k}
  + h_{(TT)ij} 
  \nonumber\\
  && \quad\quad\quad
  + \frac{2}{\alpha} K_{ij} \left(
    h_{(VL)}
    - \Delta^{-1}D^{k}\partial_{t}h_{(TV)k}
  \right)
  - \frac{2}{\alpha} K_{ij} \beta^{k} h_{(TV)k}
  \label{eq:hij-decomp-alternative-O28-09}
  , \\
  && D^{i}h_{(V)i} = 0, \quad q^{ij} h_{(TT)ij} = 0 = D^{i}h_{(TT)ij}
  \label{eq:hti-hij-decomp-conditions-O28-09}
  .
\end{eqnarray}
Here, we assume the existence of Green functions of the elliptic
derivative operators $\Delta:=D^{i}D_{i}$ and ${\cal F}$ $:=$
$\Delta$ $-$
$\frac{2}{\alpha}\left(D_{i}\alpha-\beta^{j}K_{ij}\right)D^{i}$ 
$-$
$2D^{i}\left\{\frac{1}{\alpha}\left(D_{i}\alpha-\beta^{j}K_{ij}\right)\right\}$, 
and the existence and the uniqueness of the solution $A_{i}$ to
the integro-differential equation
\begin{eqnarray}
  &&
  {\cal D}_{j}^{\;\;k}A_{k}
  + D^{m}\left[
    \frac{2}{\alpha} \tilde{K}_{mj} \left\{ \frac{}{}
      {\cal F}^{-1}D^{k}\left(
        \frac{2}{\alpha} M_{k}^{\;\;l} A_{l}
        - \partial_{t}A_{k}
      \right)
      - \beta^{k} A_{k}
      \frac{}{}
    \right\}
  \right]
  =
  L_{j}
  \label{eq:K.Nakamura-2010-note-IV-92-O28-09}
\end{eqnarray}
for given a vector field $L_{j}$.
Although the derivation of
Eq.~(\ref{eq:hti-decomp-alternative-O28-09})--(\ref{eq:hti-hij-decomp-conditions-O28-09})
is highly non-trivial, we only note that the relation
(\ref{eq:hti-decomp-alternative-O28-09})--(\ref{eq:hti-hij-decomp-conditions-O28-09})
between the variables $\{h_{ti},h_{ij}\}$ and
$\{h_{(VL)}$, $h_{(V)i}$, $h_{(L)}$, $h_{(TV)i}$, $h_{(TT)ij}\}$
is invertible if we accept above three assumptions. 
In other words, the fact that we based on these assumptions
implies that we have ignored perturbative modes which belong to
the kernel of the above derivative operators and trivial
solutions to Eq.~(\ref{eq:K.Nakamura-2010-note-IV-92-O28-09}) if
there exists. 
These modes should be separately treated in different manner.
We call these modes as {\it zero modes}.
The issue concerning about treatments of these zero modes is
called {\it zero-mode problem}, which is a remaining problem in
our general framework on higher-order general-relativistic
gauge-invariant perturbation theory.


Under these assumptions, the gauge-transformation rules for the
variables $\{h_{(VL)}$, $h_{(V)i}$, $h_{(L)}$, $h_{(TV)i}$,
$h_{(TT)ij}\}$ are summarized as follows:
\begin{eqnarray}
  &&
  {}_{{\cal Y}}h_{(VL)} - {}_{{\cal X}}h_{(VL)}
  =
  \xi_{t}
  + \Delta^{-1}D^{k}\partial_{t}\xi_{k}
  \label{eq:K.Nakamura-2010-note-B-44-O28-09}
  , \quad
  {}_{{\cal Y}}h_{(V)i} - {}_{{\cal X}}h_{(V)i}
  =
    \partial_{t}\xi_{i}
  - D_{i}\Delta^{-1}D^{k}\partial_{t}\xi_{k}
  , \\
  &&
  {}_{{\cal Y}}h_{(L)} - {}_{{\cal X}}h_{(L)}
  =
  2 D^{i}\xi_{i}
  \label{eq:K.Nakamura-2010-note-B-46-O28-09}
  , \quad
  {}_{{\cal Y}}h_{(TV)l} - {}_{{\cal X}}h_{(TV)l}
  = \xi_{l}
  ,\quad
  {}_{{\cal Y}}h_{(TT)ij} - {}_{{\cal X}}h_{(TT)ij} = 0.
\end{eqnarray}
From these gauge-transformation rules, we may define the 
components of the gauge-variant part $X_{a}$ by $X_{i}$ $:=$
$h_{(TV)i}$ and $X_{t}$ $:=$ $h_{(VL)}$ $-$
$\Delta^{-1}D^{k}\partial_{t}h_{(TV)k}$.
Then, we obtain the gauge-variant part $X_{a}$ of the
perturbation $h_{ab}$ as $X_{a}$ $:=$ $X_{t}(dt)_{a}$ $+$
$X_{i}(dx^{i})_{a}$.
Using the above variables $X_{t}$ and $X_{i}$, we can construct
gauge-invariant variables for the linear-order metric
perturbation $h_{ab}$:
\begin{eqnarray}
  \label{eq:K.Nakamura-2010-note-B-62-O28-09}
  - 2 \Phi
  &:=&
  h_{tt}
  + \frac{2}{\alpha}\left(
    \partial_{t}\alpha + \beta^{i}D_{i}\alpha - \beta^{j}\beta^{i}K_{ij}
  \right) X_{t}
  - 2 \partial_{t}X_{t}
  \nonumber\\
  &&
  + \frac{2}{\alpha} \left(
    \beta^{i}\beta^{k}\beta^{j}K_{kj} - \beta^{i}\partial_{t}\alpha 
    + \alpha q^{ij}\partial_{t}\beta_{j}
    + \alpha^{2}D^{i}\alpha
    - \alpha\beta^{k}D^{i}\beta_{k}
    - \beta^{i}\beta^{j}D_{j}\alpha
  \right)
  X_{i}
  , \\
  \label{eq:K.Nakamura-2010-note-B-64-O28-09}
  - 2 n \Psi
  &:=&
  h_{(L)} - 2 D^{i}X_{i}
  , \quad
  \nu_{i}
  :=
  h_{(V)i}
  - \partial_{t}X_{i}
  + D_{i}\Delta^{-1}D^{k}\partial_{t}X_{k}
  , \quad
  \chi_{ij} := h_{(TT)ij}.
\end{eqnarray}
Actually, we can easily confirm that these variables $\Phi$,
$\Psi$, $\nu_{i}$, and $\chi_{ij}$ are gauge invariant.
We also note that the variable $\nu_{i}$ satisfies the property
$D^{i}\nu_{i} = 0$ and the variable $\chi_{ij}$ satisfies the
properties $\chi_{ij}=\chi_{ji}$, $q^{ij}\chi_{ij}=0$, and
$D^{i}\chi_{ij}=0$.
The original components $\{h_{tt}$, $h_{ti}$, $h_{ij}\}$ of the
metric perturbation $h_{ab}$ is rewritten in terms of these
gauge-invariant variables and the variables $X_{t}$ and $X_{i}$.
These representation shows that we may define the
gauge-invariant variables ${\cal H}_{ab}$ so that  
${\cal H}_{ab}$ $:=$ $-2\Phi(dt)_{a}(dt)_{b}$ $+$
$2\nu_{i}(dt)_{(a}^{}(dx^{i})_{b)}$ $+$ 
$\left(-2\Psi q_{ij}+\chi_{ij}\right)(dx^{i})_{(a}(dx^{j})_{b)}$.
This leads to assertion of Conjecture
\ref{conjecture:decomposition-conjecture-O28-09}. $\Box$




{\it Summary ---}
We proposed an outline of a proof of Conjecture
\ref{conjecture:decomposition-conjecture-O28-09} for an arbitrary
background spacetime.
Conjecture \ref{conjecture:decomposition-conjecture-O28-09}
states that we already know the procedure to decompose the
linear-order metric perturbation $h_{ab}$ into its
gauge-invariant part ${\cal H}_{ab}$ and gauge-variant part
$X_{a}$. 
Conjecture \ref{conjecture:decomposition-conjecture-O28-09} is
the only non-trivial part when we consider the general framework
of gauge-invariant perturbation theory on an arbitrary
background spacetime.
Although there will be many approaches to prove Conjecture
\ref{conjecture:decomposition-conjecture-O28-09}, in this
article, we just proposed an outline a proof.
We also note that our arguments do not include zero modes.
The existence of zero modes is also related to the symmetry of 
the background spacetime.
To resolve this zero-mode problem, careful discussions on
domains of functions for perturbations and its boundary
conditions at $\partial\Sigma$ will be necessary. 
Besides this zero-mode problem, we have almost completed the 
general framework of the general-relativistic higher-order
gauge-invariant perturbation theory.
The outline of a proof of Conjecture
\ref{conjecture:decomposition-conjecture-O28-09} shown in this
article gives rise to the possibility of the application of our
general framework not only to cosmological
perturbations~\cite{kouchan-second-cosmo-matter-O28-09,kouchan-cosmo-second-O28-09} 
but also to perturbations of black hole spacetimes or
perturbations of general relativistic stars.
Therefore, we may say that the wide applications of our
gauge-invariant perturbation theory will be opened.
We leave these development as future works.




\begin{thebibliography}{99}
\bibitem{kouchan-gauge-inv-O28-09}
  K.~Nakamura, Prog.~Theor.~Phys. {\bf 110}, (2003), 723.
\bibitem{kouchan-second-O28-09} 
  K.~Nakamura, Prog. Theor. Phys. {\bf 113} (2005), 481.
\bibitem{kouchan-second-cosmo-matter-O28-09}
  K.~Nakamura, Phys. Rev. D {\bf 80} (2009), 124021.
\bibitem{kouchan-cosmo-second-O28-09}
  K.~Nakamura, Phys. Rev. D {\bf 74} (2006), 101301(R);
  K.~Nakamura, Prog. Theor. Phys. {\bf 117} (2007), 17;
  K.~Nakamura, Prog. Theor. Phys. {\bf 121} (2009), 1321.
\bibitem{K.Nakamura-2011-full-paper-O28-09}
  K.~Nakamura, arXiv:1105.4007 [gr-qc].
\bibitem{K.Nakamura:2010yg-O28-09}
  K.~Nakamura, Advances in Astronomy, {\bf 2010} (2010), 576273.
\end{thebibliography}
\end{document}